\RequirePackage{fix-cm}
\documentclass[smallextended]{svjour3}       
\smartqed  
\usepackage{graphicx}
\begin{document}

\title{Coupled DEM-LBM method for the free-surface simulation of heterogeneous suspensions}

\titlerunning{DEM-LBM for heterogeneous suspensions}        

\author{Alessandro Leonardi \and Falk K. Wittel \and Miller Mendoza \and Hans J. Herrmann}

\authorrunning{A. Leonardi \and F.K. Wittel \and M. Mendoza \and H.J. Herrmann} 

\institute{A. Leonardi \and Falk K. Wittel \and Miller Mendoza \and Hans J. Herrmann \at
              Institute for Building Materials, ETH Zurich\\
              Schafmattstrasse 6, CH-8093 Zurich, Switzerland\\
              Tel.: +41 44 633 04 75\\
              \email{aleonardi@ethz.ch}
}

\date{Received: date / Accepted: date}

\maketitle

\begin{abstract}
The complexity of the interactions between the constituent granular and liquid phases of a suspension requires an adequate treatment of the constituents themselves. A promising way for numerical simulations of such systems is given by hybrid computational frameworks. This is naturally done, when the Lagrangian description of particle dynamics of the granular phase finds a correspondence in the fluid description. In this work we employ extensions of the Lattice-Boltzmann Method for non-Newtonian rheology, free surfaces, and moving boundaries. The models allows for a full coupling of the phases, but in a simplified way. An experimental validation is given by an example of gravity driven flow of a particle suspension.
\end{abstract}

\section{Introduction}
\label{introduction}
Free-surface flows of heterogeneous suspensions are abundant in nature and technical applications. In principle they are multiphase materials composed of a mixture of a liquid and of solid grains of various size. A multitude of interaction mechanisms between these two phases renders the problem of their description rather difficult. For example very small grains are bounded to the liquid by electrostatic forces, while bigger ones interact mainly by viscous forces \cite{coussot1997mudflow}. Additionally, inter-grain interactions give rise to the typical complex behavior of granular matter. Often grains have a broad size distribution spanning over several orders of magnitude. Two well known examples are mixtures of mud with sand and rocks as well as suspensions of Portland cement, sand, and larger aggregates, also known as fresh concrete. While the latter is used for construction purposes \cite{Ferraris2001}, the former gives rise to devastating debris flows \cite{Iverson1997}.

The simulation of such materials is based either on continuum \cite{Nott2006,Savage1989} or on particle methods \cite{Campbell1995}, depending on whether the investigated effects arise from the physics of the fluid or the granular phase. Continuum models are appropriate when the rheological behavior of the material can be captured by rheometry techniques and phenomenological constitutive laws. However, many physical phenomena are eluded by this approach such as size and phase segregation. Examples can be found in concrete casting, where improper mixing or vibration leads to inhomogeneities in the physical properties of the hardened concrete. In debris flows, size segregation leads to locally changing flow properties. A flow front rich of large grains with high destructive power is commonly observed, followed by a fluid in a more homogeneous tail. Describing this situation by continuum methods is quite difficult or even physically inappropriate \cite{Iverson}. With particle methods, such as the Discrete 
Element Method (DEM), these phenomena can be naturally captured, which makes them an ideal tool for the study of the complex behavior of granular materials.

A complete simulation tool requires a combination of both continuum and particle description, which poses serious challenges from a computational point of view. If granular and fluid phases are fully coupled, grains represent an irregular and discontinuous boundary for the fluid domain. The relative motion of the phases complicates the picture further, because it requires the management of continuously evolving interfaces. For these reasons, traditional CFD solvers such as the standard Finite Element or Finite Volume methods, have enormous difficulties to tackle the issue. An attractive alternative is given by the Lattice-Boltzmann Method (LBM) \cite{He1997,Succi2001}, because of its extreme flexibility in the treatment of elaborate boundary conditions, its ease of implementation in parallel computing and its superior scaling when compared to traditional solvers. For these reasons, much effort has been payed to develop of a framework for particle-fluid systems combining the advantages of LBM and DEM. The early works in this field are due to Ladd  \cite{Ladd1994,Ladd1994a,Ladd2001}, who first coupled LBM and boundaries with imposed velocity. The basic model was enhanced by the use of the Immersed Boundary Method \cite{Feng2004,Owen2011,Leonardi2012a,Leonardi2012}, by the inclusion of turbulence modeling \cite{Feng2007} and extended further to the simulation of non-Newtonian rheology models \cite{Ginzburg2001,Vikhansky2008,Leonardi2011} and for free-surface flows \cite{Korner2005,Svec2012}.
Drawbacks of an approach based on the LBM are its limitation to low-Mach and relatively low-Reynolds flows, and the necessity to rely on a regular grid, since irregular grids are known to produce a complicated formalism and sometimes to lower the accuracy \cite{Ubertini2005}.

The paper is organized as follows: First a classification of particle suspensions by scales and types of physical interactions is given, before we explain how the dynamics of the different phases is addressed. In Sec. \ref{particleSolver} we summarize the DEM approach for the granular phase, followed by a section with a comprehensive description of the LBM solver for the fluid phase. Sec. \ref{extensions} explains necessary extensions to the LBM for the simulation of suspensions like fluid-particle interaction, non-Newtonian rheology or the representation of free surfaces. The experimental validation of the described model completes the manuscript in Sec. \ref{validation}, followed by a brief summary.

\section{Dynamics of Suspensions}
\label{suspensions}
The contribution of grains to the mechanics of the mixture can be of different nature depending, among other factors, on the grain size distribution. For a phenomenological classification, we use the term \textit{small scale} when electrostatic forces are dominant, \textit{medium scale} when viscous forces prevail, and \textit{large scale} when inter-particle collisional forces dominate \cite{coussot1997mudflow}. Note that the length-scales defined by grain size are by no means absolute, but depend on other parameters such as the concentration of particles, the viscosity of the liquid, and the state of the system, since the same material can exhibit different behaviors when sheared at different rates.

\textit{Small scale grain} dynamics is governed mainly by interactions of electrostatic nature, e.g. Van der Waals forces. This finer part of the grains, together with water, forms a colloidal dispersion. A complete description of this kind of material can be found elsewhere \cite{russel1992colloidal}. For practical purposes, the mechanics of colloidal dispersions is reproduced by continuum methods. A non-Newtonian model, however, is generally required, since colloidal dispersions can exhibit both shear-thinning behavior and plastic properties. In this work we choose to employ the Bingham plastic, a fluid model with a yield stress, well-known for its wide applicability \cite{Roussel2007,Whitehouse2000}. It is described as
\begin{equation}
\label{eqBingham}
   \left\{
  \begin{array}{l l}
    \dot{\gamma}=0 & \quad \textrm{if fluid does not yield}\ (\sigma<\sigma_y),\\
    \sigma = \sigma_y +\mu_{pl} \dot{\gamma} & \quad \textrm{if fluid flows}\ (\sigma>\sigma_y),\\
  \end{array} \right.\
\end{equation}
where $\dot{\gamma}$ is the magnitude of the shear rate tensor, and $\mu_{pl}$, $\sigma_y$ denote plastic viscosity and yield stress. In analogy to Newtonian fluids, an apparent viscosity (from now on, simply called viscosity) can be locally defined as the ratio of shear rate and shear stress
\begin{equation}
\label{nuBingham}
\mu_{app}=\sigma/\dot{\gamma}=\mu_{pl} + \frac{\sigma_y}{\dot{\gamma}}.
\end{equation}
As the shear rate $\dot{\gamma}$ approaches zero, the viscosity becomes infinite, giving a simple but efficient way to model plastic behavior.

\textit{Medium scale grains} are sufficiently big to elude the effects of microscopic electrostatic forces and therefore need a different numerical treatment. For them the hydrodynamic effects due to the viscous nature of the fluid become dominant. In analogy to the smaller scale, grains can be homogenized in the fluid. Obtaining an appropriate rheological behavior of the final mixture is however more difficult. When experimental data is not available, the value of the viscosity can be approximated by constitutive relations. A review of these models can be found in Ref. \cite{Stickel2005}.

\begin{figure}[t]
\centering
\includegraphics[width=106mm]{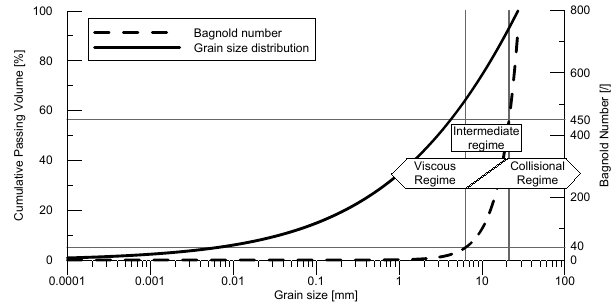}
\caption{A typical grain size distribution and its effect on the dynamics of the mixture. Big grains fall in the collisional regime, small ones in the viscous regime. A transition zone exhibits hybrid characteristics. The Bagnold number is calculated with
$\mu_f=1.0\ \textrm{Pa}\ \textrm{s}$,
$\rho = 1000\ \textrm{kg}/\textrm{m}^3$,
$\lambda = 1$,
$\dot{\gamma} = 100\ \textrm{s}^{-1}$}
\label{distribution}
\end{figure}
\textit{Large scale grain} dynamics is dominated by collisions. When collisional effects are not damped by viscosity, grains give rise to collective phenomena, such as segregation, force percolation or shock waves \cite{Herrmann1998}. Bagnold defined a dimensionless number as the ratio of grain collisional and viscous stresses \cite{Bagnold1954}. It reads
\begin{equation}
\mathrm{Ba} = \frac{\rho_s d_s^2 \lambda_s^{1/2} \dot{\gamma}}{\mu_f},
\end{equation}
where $\mu_f$ is the dynamic viscosity of the liquid, $\dot{\gamma}$ the magnitude of the shear rate, $\rho_s$ and $d_s$ denote density and characteristic diameter of the grains, and $\lambda_s$ their linear concentration (function of the solid fraction $C_s$ as $\lambda_s=1/[ (C_{s,max}/C_s)^{1/3}-1] $ with $C_{s,max}$ the maximum solid fraction). As illustrated in Fig. \ref{distribution}, the Bagnold number is used to distinguish two different regimes, where different rheological laws are observed \cite{Geographie}: Mixtures with $Ba<40$ are dominated by viscosity and therefore the shear stress grows proportionally to the shear rate. Mixtures with $Ba \geq 450$ are dominated by collisional effects and grains cannot be homogenized into a continuum description without a loss in the descriptive capabilities of the method. An intermediate range exists, where both effects are not negligible \cite{Hunt2002}.

The proposed model follows this classification to efficiently simulate and investigate suspensions. Small and medium scale grains are homogenized for a fluid formulation with a continuum  Bingham model. Large scale grains are represented by a discrete description. The advantage of this method is that only a small portion grains is explicitly represented. This fraction is representative both in terms of mass and influence on the rheology of system.

\section{Dynamics of the granular phase with the DEM}
\label{particleSolver}
The granular phase is represented by the DEM, a well-established method for granular systems \cite{Bicanic2007a}. Every grain $p$ is characterized as a Lagrangian element, with translation $\mathbf{x}_p$ and rotation $\mathbf{\phi}_p$ as degrees of freedoms. It is subjected to multiple interactions that lead to a resultant force $\mathbf{F}_p$ and moment $\mathbf{M}_p$. These interactions can either be due to collisions, hydrodynamics or volumetric forces and are functions of position, orientation and velocity of the particles:
$\mathbf{F}_p=\mathbf{F}_p\left(\mathbf{x}_p,\dot{\mathbf{x}}_p,
\mathbf{\phi}_p,\dot{\mathbf{\phi}}_p\right)$,
$\mathbf{M}_p=\mathbf{M}_p\left(\mathbf{x}_p,\dot{\mathbf{x}}_p,
\mathbf{\phi}_p,\dot{\mathbf{\phi}}_p\right)$.

In the simplest case, DEM particles have spherical shape allowing for fast contact detection and calculation of the overlap $\xi_{p,q}$ between particles $p$ and $q$ namely
\begin{equation}
\label{contact}
 \xi_{p,q}=||\mathbf{d}_{p,q}||-R_p-R_q.
\end{equation}
Here $\mathbf{d}_{p,q}$ denotes the distance between the center of the spheres with radii $R_p$, $R_q$. Unfortunately, for most practical applications spheres can only be used as a first approximation of the real particle shape. Note that spheres exhibit artificial mixing and rolling behavior, which is absent in natural system that are not composed of spheres. To overcome these effects we use composite elements, created by aggregating a set of spherical particles. While preserving the simplicity of the contact calculation, composite elements allow for a more realistic representation of granular effects, in particular in the limit of dense concentrations.

Particle-particle interactions are written as the outcome of collisional events between particles. Although particles are geometrically described as rigid spheres, the overlap  $\xi_{p,q}$ between particles $p$ and $q$ is used to calculate collisional forces and to represent the elastic deformation. We use the law for elastic spheres,
\begin{equation}
\label{hertz}
 \mathbf{F}^n_{p,q}=\frac{2}{3}\frac{Y\sqrt{R_{eff}}}{\left(1-\nu^2\right)}
 \left(\xi_{p,q}^{3/2}+A\sqrt{\xi_{p,q}}\frac{d\xi_{p,q}}{dt}\right)\mathbf{n}_{p,q},
\end{equation}
where $Y$ and $\nu$ are the Young modulus and the Poisson's ratio of the material, $A$ is a damping constant \cite{Brilliantov1996}, $R_{eff}$ the effective radius defined as $R_{eff}=R_p R_q / (R_p+R_q)$ and $\mathbf{n}_{p,q}$ the normal vector of the contact surface.
The tangential contact force is considered to be proportional to the component of the relative velocity of the two spheres
laying on the contact surface $\mathbf{u}^t_{rel}$ as
\begin{equation}
 \mathbf{F}^t_{p,q}=-\textrm{sign}\left(\mathbf{u}^t_{rel}\right)\cdot
 \textrm{min}\left(\eta||\mathbf{u}^t_{rel}||, \mu_d ||\mathbf{F}^n_{p,q}|| \right)
 \mathbf{t}_{p,q},
\end{equation}
with the tangential shear viscosity coefficient $\eta$ and the dynamic friction coefficient $\mu_d$, thus including Coulomb friction. The tangential unit vector $\mathbf{t}_{p,q}$ is obtained normalizing the tangential relative velocity. Wall contacts are calculated in a similar fashion.

The time evolution of the system is solved by integrating Newton's second law,
\begin{eqnarray}
m_p\ddot{\mathbf{x}}_p =
\mathbf{F}_p, &
\mathbf{J}_p\ddot{\mathbf{\phi}}_p =
\mathbf{M}_p -
\dot{\mathbf{\phi}}_p \times \left(\mathbf{J}_p\dot{\mathbf{\phi}}_p \right).
\end{eqnarray}
While the translational motion is naturally solved in system coordinates, the rotational motion requires additional considerations. We use quaternion algebra rather that Euler angles to represent the orientation of the elements, and calculate and invert rotation matrices without singularities \cite{Carmona2008}. Newton's equations in the body-fixed reference frame produce $6\times N$ scalar equation, where $N$ is the number of elements. The Gear predictor-corrector differential scheme is used to integrate them \cite{Gear1971}. During the predictor step, tentative values for particle position, orientation and their derivatives are computed, using a Taylor expansion of the previous time step values. The predicted values are then used to check for contacts, compute collisional forces and solve Newton's equations of motion. For the corrector step, the difference between the predicted values for acceleration and their counterpart resulting from Newton's equation is computed. This difference is used to calculate 
the new corrected values for position, orientation and their derivatives.

The DEM time step should be small enough to resolve the particle contacts. If $t_c$ is the collision time, then ${\Delta t}^{\textrm{DEM}}\ll t_c$, usually ${\Delta t}^{\textrm{DEM}}< 0.1 t_c$. The collision time can be estimated for a Hertzian contact as
\begin{eqnarray}
 t_c=2.5\left(\frac{m_{eff}^2}{k^2||\mathbf{u}^n_{rel}||}\right)^{\frac{1}{5}} \textrm{, with}
 & k=\frac{8}{15}\frac{Y}{1-\nu^2}\sqrt{R_{eff}},
\end{eqnarray}
where $m_{eff} = m_p m_q /\left(m_p+m_q\right)$ is the effective mass and $\mathbf{u}^n_{rel}$ denotes the normal relative velocity at the contact point at the beginning of the collision.

\section{Fluid dynamics with the LBM}
\label{fluidSolver}
\begin{figure}[t] \centering \includegraphics[width=85mm]{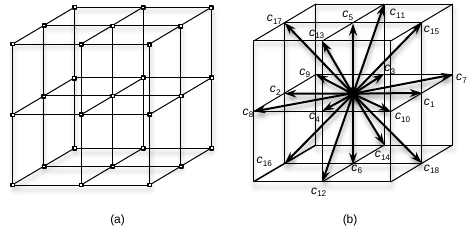}
  \caption{a) The regular space discretization of the lattice. b) The
    19 discrete velocities allowed in the D3Q19 lattice
    configuration} \label{lattice} \end{figure}
The DEM described in the previous section is coupled with the LBM for
solving the fluid phase. LBM has been evolving very fast in the last
two decades and is considered to be one of the most attractive alternatives
to traditional CFD solvers, especially when problems feature
complex boundary conditions. It originates from the Boltzmann kinetic
theory for the evolution of molecular systems \cite{Benzia1992,Chen1998}. The fluid is
described using a distribution function, $f(\mathbf{x},\mathbf{c},t)$,
defined as the probability density of finding molecules with velocity
$\mathbf{c}$ at a location $\mathbf{x}$ and at a given time $t$. 

In the LBM, the velocity space is discretized by a finite number of
velocity vectors, $\mathbf{c}_i$, such that $f_i(\mathbf{x},t) \equiv
f(\mathbf{x},\mathbf{c}_i,t)$. We choose to employ the D3Q19 lattice
cell configuration ($3$ dimensions and $19$ velocities, see Fig.
\ref{lattice}), which provides the required symmetries to correctly
recover the incompressible Navier-Stokes equations. In this work, for
simplicity, we use dimensionless lattice units ($\delta_{x,y,z}=1,\
\delta_t =1$).

The reconstruction of macroscopic physical variables such as density
$\rho_f$ and velocity $\mathbf{u}_f$ can be done at every location
$\mathbf{x}$ and time $t$ by computing the first two moments of the
distribution function $f_i(\mathbf{x},t)$
\begin{eqnarray}
\label{densityVelocity}
 \rho_f(\mathbf{x},t)=\sum\limits_i f_i(\mathbf{x},t), & 
 \mathbf{u}_f=\sum\limits_i f_i(\mathbf{x},t)\mathbf{c}_i/\rho_f(\mathbf{x},t).
\end{eqnarray}
The distribution function evolves according to the Lattice Boltzmann
Equation (LBE), which is written
\begin{eqnarray}
\label{newfunctions}
f_i (\mathbf{x}+\mathbf{c}_i,t+1)= f_i (\mathbf{x},t)+\Omega_i (\mathbf{x},t) .
\end{eqnarray}
$\Omega_i$ represents the collision operator, which in our case
corresponds to the linear approximation given by Bhatnagar-Gross-Krook
\cite{Bhatnagar1954},
\begin{eqnarray} \label{collision} 
\Omega_i
(\mathbf{x},t)=\frac{f_i^{eq} (\mathbf{u}_f,\rho_f)-f_i
  (\mathbf{x},t)}{\tau(\mathbf{x},t)} , 
\end{eqnarray}
where $\tau$ is the relaxation time and $f_i^{eq}$ is the equilibrium
distribution function. The relaxation time is directly related to the
viscosity of the fluid
\begin{eqnarray}
\label{nu}
\mu_f(\mathbf{x},t)=\frac{\tau(\mathbf{x},t)-1/2}{3}.
\end{eqnarray}
For a Newtonian fluid, the relaxation time $\tau$ is a constant and
global parameter. However, as stated in Sec. \ref{suspensions}, in
order to represent most suspensions, a non-Newtonian formulation
should be employed. To do this, the relaxation time is treated as a
local variable $\tau(\mathbf{x},t)$. The equilibrium distribution function $f_i^{eq}$ is an expansion in
Hermite polynomials of the Maxwell-Boltzmann distribution in the limit
of small velocities \cite{Shan1998}. Using the local macroscopic velocity
$\mathbf{u}_f$ and density $\rho_f$, this yields
\begin{eqnarray}
\label{equilibrium}
f_i^{eq} (\mathbf{u}_f,\rho_f) = \rho_f w_i \left(1+3\mathbf{c}_i\cdot\mathbf{u}_f +\frac{9}{2}
\left(\mathbf{c}_i\cdot\mathbf{u}_f\right)^2 -\frac{3}{2}\mathbf{u}_f\cdot\mathbf{u}_f\right),
\end{eqnarray}
where the weights $w_i$ are constants that ensure the recovering of
the first and second moments of the distribution function (Eq.
\ref{densityVelocity}). For the D3Q19 lattice configuration they are
\begin{eqnarray}
w_i = \left\{
  \begin{array}{l l}
    1/3 & \quad \textrm{for}\ i=1\\
    1/18 & \quad \textrm{for}\ i=2\textrm{...}7\\
    1/36 & \quad \textrm{for}\ i=8\textrm{...}19.\\
  \end{array} \right.\
\end{eqnarray}

To introduce an external force, we employ the scheme developed by Guo
et al. \cite{Guo2002}, which consists in modifying Eq.
\ref{newfunctions} as
\begin{equation}
  \label{forceFunctions}
  f_i (\mathbf{x}+\mathbf{c}_i,t+1)= f_i (\mathbf{x},t)+\Omega_i (\mathbf{x},t)+F_i (\mathbf{x},t),
\end{equation}
where $F_i (\mathbf{x},t)$ is an additional distribution function due
to the force field $\mathbf{F}$, which can be calculated in a similar
fashion as the equilibrium distribution,
\begin{equation}
 F_i (\mathbf{x},t)=w_i\left(1-\frac{1}{2\tau}\right)
 \left[3\left(\mathbf{c}_i-\mathbf{u}_f\right)+
 9\mathbf{c}_i\left(\mathbf{c}_i\cdot\mathbf{u}_f\right)\right]\mathbf{F}.
\end{equation}
With this technique, the computation of the macroscopic velocity field in Eq. \ref{densityVelocity}
also needs to be modified,
\begin{eqnarray}
 \mathbf{u}_f=\left(\sum\limits_i f_i(\mathbf{x},t)\mathbf{c}_i
 +\mathbf{F}/2\right)/\rho_f(\mathbf{x},t).
\end{eqnarray}

The described approach reproduces the Navier-Stokes equations in the
incompressible limit. The pressure is directly computable from the
density as
\begin{equation}
  P_f(\mathbf{x},t)=\rho_f(\mathbf{x},t)\cdot c_s^2,
\end{equation}
where $c_s$ is the speed of sound of the fluid, which corresponds to
$c_s=1/\sqrt{3}$ (lattice units). The stability and accuracy of the LBM are guaranteed for small Mach
numbers, $Ma \equiv ||\mathbf{u}||/c_s \ll 1$.

For every time step one first calculates the macroscopic variables,
using Eq. \ref{densityVelocity}, and the corresponding
equilibrium distribution, from Eq. \ref{equilibrium}. Then, one
uses Eq. \ref{newfunctions} to evolve the distribution function,
which provides the new density and velocity of the fluid for the next
time step. Being solved mostly at a local level, the scheme can be
easily implemented in a parallel environment \cite{Monitzer2012}.

\section{Extensions of the LBM for the simulation of suspensions}
\label{extensions}
To widen the range of applicability of the model to heterogeneous suspensions, we need to incorporate a
few more features. First, we introduce no-slip moving boundaries,
necessary for the coupling with the DEM, and second, in Sec.
\ref{freeSurface}, we extend the model to simulate free surfaces.
Finally, Sec. \ref{rheologyModel} describes the method for
non-Newtonian formulations.

\subsection{Coupling with particles}
\label{boundaryConditions}
\begin{figure}[t] \centering \includegraphics[]{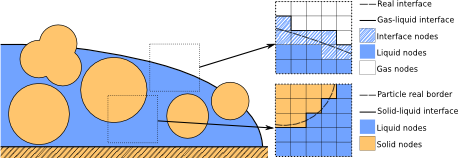}
  \caption{Sketch showing how particles or solid objects
    are discretized on the regular lattice. The free-surface is
    treated in a similar way, with a special type of nodes defining
    the interface}
\label{sketch}
\end{figure}
The coupling with the DEM and the treatment of no-slip boundary
conditions are performed at a local level by modifying the LBE.
Lattice nodes are divided into fluid and solid nodes, the latter
ones representing particles and walls (see Fig. \ref{sketch}).
Solid nodes are inactive, i.e. on them the LBE is not
solved. No-slip is performed with the so-called \emph{bounce-back
  rule}: every time a distribution function $f_{i} (\mathbf{x},t)$ is
streaming in the direction $i$ towards a solid node, it gets reflected
back in the opposite direction $i'$. If the boundary is moving, the
reflected distribution needs to be corrected as
\begin{eqnarray}
\label{SBB}
 f_{i'} (\mathbf{x},t+1) = f_{i} (\mathbf{x},t)
 - 6 w_i \rho \mathbf{u}_w \cdot \mathbf{c}_i,
\end{eqnarray}
where $\mathbf{u}_w$ is the local velocity of the wall at the
bounce-back location. If the wall represents the surface of a
particle, the local velocity can be obtained as
\begin{equation}
 \mathbf{u}_w=\mathbf{u}_p + \mathbf{r}_w \times \mathbf{\omega}_p,
\end{equation}
where $\mathbf{u}_p$ and $\mathbf{\omega}_p$ are the linear and
angular velocity of the particle, and $\mathbf{r}_w$ is the vector
connecting its center of mass with the bounce-back location. The
momentum exchange experienced by the reflected distribution can also
be used to compute the force exerted on the wall when integrated
over all bounce-back locations,
\begin{eqnarray}
\label{hydroforce}
 \mathbf{F}_p=\sum \left( 2 f_{i} (\mathbf{x},t)
 - 6 w_i \rho_f \mathbf{u}_w \cdot \mathbf{c}_i \right) \mathbf{c}_i.
\end{eqnarray}
Solid boundaries treated this way are located halfway between solid
and active nodes. This technique was developed for moving boundaries
by Ladd \cite{Ladd1994} and Aidun and Lu \cite{Aidun1995}.

Particles move over a fixed,
regular grid. Of course the node classification into fluid and solid is not fixed but needs to be
updated. Following the particle motion, fluid
nodes are created (deleted) in the wake (front) of moving particles.
The macroscopic density and the velocity of the newly created nodes are
calculated as the average over the values in the neighborhood
as initial values for the distribution function $f_i (\mathbf{x},t)$ through Eq.
\ref{equilibrium}. Deleted fluid nodes are converted to solid ones and
therefore made inactive. Both processes introduce small variations in
the global mass and momentum. However, due to the fact that all our
simulations are performed in the incompressible limit (variation of
density are very small), and that fluid nodes close to a particle
possess nearly the same velocity as the particle, we expect these variations to be
negligible. Another problem is the representation of the particle boundaries on the regular lattice,
which leads to a zig-zag approximation of the spherical shapes. An alternative way to overcome these problems is the use of the Immersed Boundary Method \cite{Feng2004} or of a fictitious domain \cite{Glowinski1999,Glowinski2001}.
Both methods are more precise and smooth the ill effects of
particles traveling though the lattice. At the same time, they require additional computations and are therefore avoided following the spirit of this paper.

When two particles approach each other, the distance between the
surfaces can become smaller than the lattice node spacing, resulting
in an imprecise resolution of the collision process. To overcame
this problem, we use the lubrication theory of Nguyen and Ladd
\cite{Nguyen2002}. In this theory, when two particle are moving with a
relative velocity $\mathbf{u}_{rel}$, the correction force
\begin{equation}
\mathbf{F}^{lub}_{p,q}=-6\mu_f||\mathbf{u}^n_{rel}||
 R_{eff}^2\left(1/s_{p,q}-1/d_{lub}\right)\mathbf{n}_{p,q},
\end{equation}
is added, where $s_{p,q}=-\xi_{p,q}$ is the distance between the particle surfaces and
$d_{lub}$ denotes a cut-off distance above which no force is computed.

\subsection{Free surface representation}
\label{freeSurface}
We employ the mass tracking algorithm described in Refs. \cite{Chen1998,Mendoza2010} which, despite
its simplicity, leads to a stable and accurate surface evolution.
Fluid nodes are further
divided into liquid, interface and gas nodes: Liquid and interface
nodes are considered active, and the LBE is solved. The remaining nodes are the gas nodes
and are inactive, with no evolution equation. Liquid and gas nodes are
never directly connected, but through an interface node (see Fig. \ref{sketch}).

An additional macroscopic variable for the mass
$m_f(\mathbf{x},t)$ stored in a node is required, defined as
\begin{eqnarray}
\left\{
  \begin{array}{l l}
    m_f(\mathbf{x},t)=\rho_f(\mathbf{x},t) & \quad \textrm{if the node is liquid,}\\
    0<m_f(\mathbf{x},t)<\rho_f(\mathbf{x},t) & \quad \textrm{if the node is interface,}\\
    m_f(\mathbf{x},t)=0 & \quad \textrm{if the node is gas.}\\
  \end{array} \right.\
\end{eqnarray}
The mass is updated using the equation
\begin{eqnarray}
 m_f (\mathbf{x},t+1) = m_f (\mathbf{x},t) + \sum_i \alpha_i \left[ f_{i'} (\mathbf{x}+\mathbf{c}_i,t)
   - f_{i} (\mathbf{x},t)\right] ,
\end{eqnarray}
where $\alpha_i$ is a parameter determined by the nature of the neighbor node in the $i$ direction,
\begin{eqnarray}
  \alpha_i = \left\{
  \begin{array}{l l}
    \frac{1}{2}\left[m_f (\mathbf{x},t) + m_f (\mathbf{x}+\mathbf{c}_i,t)\right] & \quad \textrm{if the neighbor node is interface,}\\
    1 & \quad \textrm{if the neighbor node is liquid,}\\
    0 & \quad \textrm{if the neighbor node is gas.}\\
  \end{array} \right.\
\end{eqnarray}
When the mass becomes zero ($m_f (\mathbf{x},t) = 0$), the interface
node is transformed into gas, with all liquid nodes connected to it
becoming interface. Analogously, an interface node whose mass reaches
the density ($m_f (\mathbf{x},t) = \rho_f (\mathbf{x},t)$) is transformed
into liquid, and all connected gas nodes become interface. However,
due to the discrete integration, these equalities are not in general
satisfied. The surplus of mass is equally distributed to the
neighboring interface nodes, conserving the total mass of the system.

Because gas nodes are not active, there are no distribution functions
streaming from gas nodes to interface nodes. These missing
distribution functions are computed from the macroscopic variables at
the interface, atmospheric density $\rho_{atm}$ and interface velocity
$\mathbf{u}_{int}$, as
\begin{eqnarray}
 f_{i'} (\mathbf{x}+\mathbf{c}_{i'},t+1) =
 f^{eq}_{i} (\mathbf{u}_{int},\rho_{atm})
 +f^{eq}_{i'} (\mathbf{u}_{int},\rho_{atm})
- f_{i} (\mathbf{x},t) .
\end{eqnarray}
Note that this implies that gas nodes have the same macroscopic
velocity as the connected interface nodes.

\subsection{Bingham plastic rheology model}
\label{rheologyModel}
\begin{figure}[t]
\centering
\includegraphics[width=0.5\textwidth]{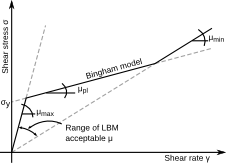}
\caption{Representation of the rheology model employed for plastic fluids. The approximation of the Bingham model is limited by the maximum and minimum acceptable values for the relaxation time $\tau$ and therefore for the viscosity $\mu_f$}
\label{trilinear}
\end{figure}
The presence of the small particle fraction in the fluid leads to non-Newtonian behavior, that needs to be considered. For the LBM this implies that the relaxation time $\tau$ is not a global parameter for the system, but rather
$\tau=\tau(\mathbf{x},t)$. A non-linear dependency of viscosity, and thus of $\tau$, on
the shear rate requires an explicit computation of the shear rate
tensor. This can be done with ease in the LBM from the non-equilibrium
part of the distribution functions,
\begin{eqnarray}
 \dot{\gamma}_{ab} (\mathbf{x},t) = \frac{3}{2\tau(\mathbf{x},t)}
 \sum_i \mathbf{c}_{i,a} \mathbf{c}_{i,b}
 \left( f_i (\mathbf{x},t) -f_i^{eq}(\mathbf{x},t)\right).
\end{eqnarray}
With the second invariant of the shear rate tensor
\begin{equation}
  \Gamma_{\dot{\gamma}}(\mathbf{x},t)=
  \sum_a \sum_b  \dot{\gamma}_{ab}\dot{\gamma}_{ab},
\end{equation}
the magnitude of the shear rate is calculated as
\begin{equation}
  \dot{\gamma}(\mathbf{x},t)=\sqrt{2 \Gamma_{\dot{\gamma}}(\mathbf{x},t)}.
\end{equation}
This can be included in any constitutive equation for purely viscous
fluids. As outlined before (Sec. \ref{suspensions}), we choose
the Bingham constitutive model and get a new form of Eq. \ref{nu} for the explicit update of $\tau$,
\begin{eqnarray}
\label{const}
\tau(\mathbf{x},t)=\frac{1}{2}+3\left(\mu_{pl} + \frac{\sigma_y}{\dot{\gamma}(\mathbf{x},t)}\right).
\end{eqnarray}

The accuracy and stability of LBM are guaranteed only over a
certain range of values for $\tau$. This limits the applicability of
Eq. \ref{const}, because $\tau$ diverges when
$\dot{\gamma}\rightarrow 0$. Following \v{S}vec et al.
\cite{Svec2012}, we use a simple solution to this problem, imposing
that $\tau_{min} \leq \tau(\mathbf{x},t) \leq \tau_{max}$.
Reasonable values for $\tau_{min}$ and $\tau_{max}$ are,
respectively, $0.501$ and $3.5$. The constitutive equation
arising from this approach is that of a tri-viscosity fluid (see Fig. \ref{trilinear}). If
$\mu_{f,min}<\mu_{pl}$ the model represents a bi-viscosity, and if
$\mu_{f,max} \gg \mu_{pl}$, the approximation of the Bingham model is
fair. With these extensions the model is complete and we can address examples.

\section{Experimental validation by a gravity-driven flow}
\label{validation}
\begin{figure}[t]
\centering
\includegraphics[width=94mm]{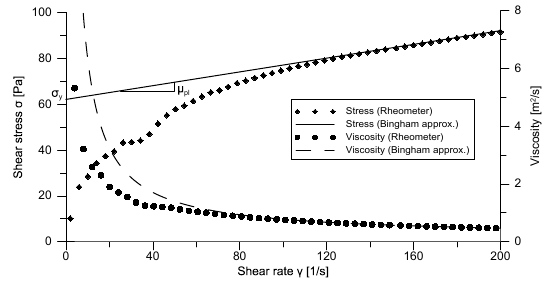}
\caption{Rheology test on the fresh cement paste. A linear Bingham approximation is used to fit the data, obtaining $\mu_{pl}$ and $\sigma_y$}
\label{bingham_graph}
\end{figure}
The capabilities of the model are shown by comparing with an experiment, featuring a free-surface flow of a suspension under the effect of gravity. We employ fresh concrete, since it poses all the challenges necessary to validate the method: a non-Newtonian rheology and an irregular granular phase. The cement paste is obtained with a commercial Portland cement of type CEM I $42.5 \textrm{N}$. Water is added until a water/cement ratio of $0.4$ is reached. The rheology of the obtained paste is measured with a coaxial rotational viscometer Haake RV20. The measurement procedure consists in the uniform shearing of the paste at $200\ \textrm{s}^{-1}$ for $120\ \textrm{s}$, followed by a shear rate continuous ramp from $0$ up to $200\ \textrm{s}^{-1}$ occurring over $120\ \textrm{s}$  \cite{Geiker2002}. The obtained rheological curve is shown in Fig. \ref{bingham_graph}.

\begin{figure}[t]
\centering
\includegraphics[width=0.7\textwidth]{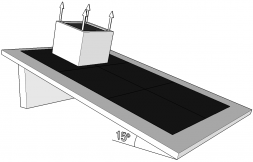}
\caption{Experiment setup with an inclined plane of $933 \times 700 \textrm{mm}$. The internal size of the box is $150\times 150\ \textrm{mm}$. The dark gray area represents the surface covered by the sandpaper}
\label{geometry}
\end{figure}

\begin{figure}[t]
\centering
\includegraphics[width=100mm]{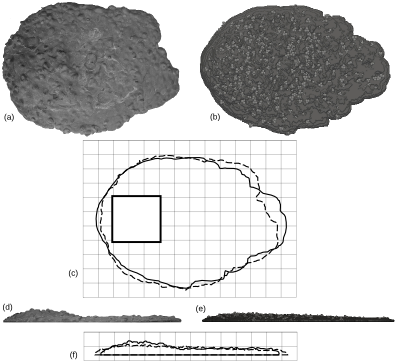}
\caption{Final shape of the flowing mass. a,d) Numerical shape; b,e) Experimental shape. Fluid mass opacity in the numerical shape is lowered for a better visualization of particles; c,f) Comparison of numerical shape (solid line) and experimental shape (dashed line).
The background grid has $5 \textrm{cm}$ spacing}
\label{shapes}
\end{figure}

The paste is mixed with $1000$ silica rounded pebbles with radius $R=4.0\div 8.0\ \textrm{mm}$. The total weight of the grains is $2.629\ \textrm{kg}$, and the density $\rho_s = 2680\ \textrm{kg}/\textrm{m}^3$. The components are mixed in a bowl until homogenization and then vibrated for degassing. The final mixture is poured in a $150\times 150\ \textrm{mm}$ rectangular box, open on top and bottom and positioned over a wooden board inclined at $15^{\circ}$. The board surface is upholstered with sandpaper and wetted before the start of the test. The test is performed by steadily lifting the box, and letting the sample spread on the board under the sole effect of gravity. The flow falls in the intermediate regime (see Fig. \ref{distribution}). Collisional effects are therefore not dominant, but still important. The geometry of the test is illustrated in Fig. \ref{geometry}, and Fig. \ref{shapes} (a) is a picture of the final deposition of the sample.

\begin{figure}[t]
\centering
\includegraphics[width=87mm]{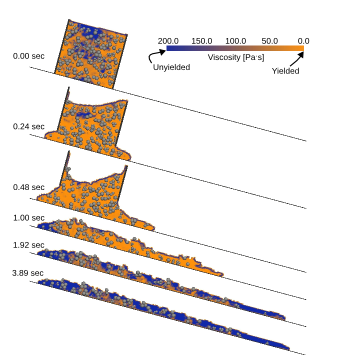}
\caption{Dynamic viscosity contour on the longitudinal cross section of the simulation. Particles are represented in light gray and walls in dark gray. The yielded region of the fluid grows from a small portion close to the box walls to the whole sample during the first part of the simulation. When a new equilibrium is reached, the yielded region reduces and the flow is slowed}
\label{viscosity}
\end{figure}

The same environment is set up with a simulation on an LBM lattice of $350 \times 250 \times 80$ nodes, with the lattice spacing corresponding to $2.0\cdot 10^{-3} \textrm{m}$ in physical units. The initial configuration of the fluid is a cube with edge length of $0.15 \textrm{m}$, corresponding to $75 \times 75 \times 75$ liquid nodes. The pebbles are represented with $1000$ discrete elements, each composed of 4 spheres with tetrahedral structure. The total number of spheres is $4000$. The box is represented by a set of moving walls and is set as solid boundary both for fluid and granular solvers. The lifting speed of the box is $0.15 \textrm{m}/\textrm{s}$. The properties of the fluid are obtained from the viscometer data, as represented in Fig. \ref{bingham_graph}. A good fit is obtained with a Bingham model with plastic viscosity $\mu_{pl}=0.15\ \textrm{Pa}\cdot \textrm{s}$ and yield stress $\sigma_y=62\ \textrm{Pa}$. The model is imprecise for lower shear rates, which is one of the limitations of the 
chosen linear approach. Fig. \ref{viscosity} shows the results of the simulation on the longitudinal cross section of the sample. The evolution of the shear rate and the particle distribution can be tracked continuously.
The final shapes of the experimental and numerical solution are compared in Fig. \ref{shapes} (c), showing excellent agreement.

A good compromise between stability and speed is obtained with a time step of ${\Delta t}^{\textrm{LBM}}=3.0\cdot 10^{-5}\ \textrm{s}$.
This sets the maximum allowable speed in the system as $0.667\ \textrm{m}/\textrm{s}$. The parameters in lattice units are then viscosity $\mu^{\textrm{LBM}}_{pl}=6.25\cdot10^{-4}$, relaxation time $\tau^{\textrm{LBM}}_y=7.75\cdot 10^{-6}$, and gravity $||\mathbf{g}^{\textrm{LBM}}||=4.41\cdot 10^{-6}$. The simulation is stopped when $95\%$ of the fluid has reached the maximum viscosity. The total simulation time is 63 hours, with a parallel run on 4 cores with an Intel Xeon E5-1620 processor at 3.60 GHz..

\section{Summary}
\label{summary}
In this paper a model for the simulation of the flow of suspensions was proposed. The multiscale nature of the model is justified by the different interaction mechanisms acting between the liquid and the granular phase. A practical mean of phenomenological classification of interactions is given by the Bagnold number: Small grains are considered to be governed by the viscous nature of the liquid and are modeled as part of the fluid phase itself with the use of a plastic non-Newtonian formulation. Grains with a sufficiently large size are dominated by collisional mechanisms. This is modeled with a two-way coupling between fluid and grains, along with the resolution of particle contacts.

The problem was solved with a hybrid of the Discrete Element Method for grains and the Lattice-Boltzmann Method for fluids. A combination of the most successful advances in these methods was employed. The mass-tracking algorithm allows an inexpensive way to simulate free surfaces, while the variable relaxation time formulation can reproduce non-Newtonian constitutive laws. The hydrodynamic interactions with the granular phase were fully solved with the bounce-back rule for coupling non-slip moving boundaries and fluid. The proposed model finds its best application in the simulation of real flows and in particular of heterogeneous suspensions with a granular phase that features a complete size distribution, due to its multiscale nature. The intrinsic advantages of the Lattice-Boltzmann solver, with its high-
level performance and its relatively simple implementation make it a good choice for the fast development of such methods. Moreover, the core of the solver works at a local level, making the parallelization of the code easy and natural.
Grain-grain interactions were solved with a Discrete Element Method. We assured that the scaling of the particle solver was not too far from the almost linear performances of the fluid solver. The Hertzian contact law was used, and a formulation for non-spherical particles was included.
The capabilities of the approach were shown by comparing to an experimental free-surface flow of a fresh concrete sample. An excellent agreement between numerical and experimental data was found in the comparison of the final shape of the sample. The results of the simulation can provide insight into the mechanics of the flow. The spatial distribution of particles can be tracked, along with the variables of the flow: velocity, pressure, shear rate and viscosity.

Another challenging application of the model is the prediction of debris flows, which can hardly be assessed experimentally. Future works will focus on the rheology of debris materials and on the full simulation of events for deeper physical understanding, and on techniques for the design of effective protection measures.

\begin{acknowledgements}
The research leading to these results has received funding from the European Union (FP7/2007-2013) under Grant Agreement No. 289911, as well as from the European Research Council (ERC) through Advanced Grant No. 319968-FlowCCS. The authors are grateful for the support of the European research network MUMOLADE (Multiscale Modelling of Landslides and Debris Flows).
\end{acknowledgements}


\end{document}